\documentclass[12pt,showpacs,preprintnumbers,amsmath,amssymb]{revtex4}
\usepackage{graphicx}% Include figure files
\usepackage{color} %example: {\color{red} text}

\begin{document}

\newcommand{\pderiv}[2]{\frac{\partial #1}{\partial #2}}
\newcommand{\deriv}[2]{\frac{d #1}{d #2}}
\newcommand{\eq}[1]{Eq.~(\ref{#1})}  %example: \eq{eq:NFPE}
\newcommand{\infint}{\int \limits_{-\infty}^{\infty}}

\title{Generalized Nonlinear Proca Equation and Its 
Free-Particle Solutions}

\vskip \baselineskip

\author{F.~D. Nobre$^{1}$\footnote{E-mail address: fdnobre@cbpf.br }
and A.~R. Plastino$^{2}$\footnote{
% Corresponding author: 
E-mail address: arplastino@unnoba.edu.ar} }
\affiliation{
$^{1}$Centro Brasileiro de Pesquisas F\'{\i}sicas and \\
National Institute of Science and Technology for Complex Systems, \\
Rua Xavier Sigaud 150,
22290-180, Rio de Janeiro - RJ, Brazil \\
$^{2}$CeBio y Secretar\'{\i}a de Investigaci\'on, 
Universidad Nacional Buenos Aires-Noreoeste, 
UNNOBA-Conicet, Roque Saenz Pe\~na 456, Junin, Argentina}

%\date{\today}

\newpage
%\vskip \baselineskip

\begin{abstract}
We introduce a non-linear extension of Proca's field theory for massive vector (spin $1$)
bosons. The associated relativistic nonlinear wave equation
is related to recently advanced nonlinear extensions of the Schroedinger, 
Dirac, and Klein-Gordon equations inspired on the
non-extensive generalized thermostatistics. This is a theoretical framework
 that has been applied in recent years to several problems
 in nuclear and particle physics, gravitational physics, 
 and quantum field theory. The nonlinear Proca equation
 investigated here has a power-law nonlinearity characterized
 by a real parameter $q$ (formally corresponding 
 to the Tsallis entropic parameter) 
% An exact classical field theory has been applied recently to
% generalized nonlinear versions of Schroedinger and Klein-Gordon equations.
%These generalizations were achieved by introducing nonlinear terms,
%characterized by exponents depending on an index $q$, 
in such a way that the standard linear Proca wave equation is 
recovered in the limit $q \rightarrow 1$. We derive the nonlinear Proca 
equation from a Lagrangian that, besides the usual vectorial field $\Psi^{\mu}(\vec{x},t)$, 
involves an additional field $\Phi^{\mu}(\vec{x},t)$.
% The vectorial field $\Phi^{\mu}(\vec{x},t)$, which is
% shown to obey an additional nonlinear evolution equation, becomes
% $\Psi^{\mu*}(\vec{x},t)$ only when $q \rightarrow 1$.
We obtain exact time dependent soliton-like solutions for these fields
having the form of a $q$-plane wave, and show that both field 
equations lead to the relativistic energy-momentum relation
$E^{2} = p^{2}c^{2} + m^{2}c^{4}$ for all values of $q$. This suggests that the present nonlinear theory 
constitutes a new field theoretical representation of particle dynamics.
In the limit of massless particles the present $q$-generalized Proca theory reduces 
to Maxwell electromagnetism, and the $q$-plane
waves yield localized, transverse solutions of Maxwell equations.
Physical consequences and possible applications are discussed.

\vskip 0.2cm

%\vskip \baselineskip
%\vspace{2cm}
%\medskip


\end{abstract}
\maketitle

%\newpage

\section{Introduction}

The aim of the present contribution is to advance and explore some features of a non-linear
 extension of Proca's field equation. This proposal is motivated by recent developments concerning
 nonlinear extensions of the Schroedinger, Dirac, and Klein-Gordon equations 
 \cite{NRT2011,NRT2012,PT2013,CPP2013,PSNT2014,ARSML2015,RN2013} related to the
 non-extensive generalized thermostatistics \cite{T2009,T88}. 
 These nonlinear equations are closely related to a family of power-law
 nonlinear Fokker-Planck equations that describe the spatio-temporal 
 behavior of various physical systems  and processes and have been studied intensively
 in recent years \cite{F2005,PP95,TB96,SNC2007a,ASMNC2010,RNC2012,NSC2012,CSNA2014}.

The Proca equation \cite{P_1936} constitutes, along with the Dirac and the Klein-Gordon 
equations, one of the fundamental relativistic wave equations \cite{gregre}. 
It describes massive vector (spin $1$) boson fields. Historically it played an 
important role due, among other things, to its relation with Yukawa's work on
mesons \cite{Pauli_1941}. Proca's equation shields 
a generalization of Maxwell's electromagnetic field theory, by incorporating 
the effects of a finite rest mass for the photon \cite{J_1998}. As such,
it is a basic ingredient of the theoretical framework for experimental studies aiming at the 
determination of upper bounds for the mass of the photon \cite{GN2010,TLG_2005}. 
This is a basic line of enquiry
that can be regarded as having its origins in experimental work by Cavendish and Maxwell, 
with their explorations of possible deviations from Coulomb's electrostatic force law \cite{TLG_2005}  
(although they, obviously, did not formulate this problem in terms of the mass of the photon). 
Within the modern approach, the Proca field equation allows for definite quantitative predictions 
concerning diverse physical effects originating in a finite mass of the photon (including the 
aforementioned deviations from Coulomb's law). This, in turn, motivates concrete experiments 
for the search of the mentioned effects, and for determining concomitant upper bounds for the 
photon's rest mass. Besides the problem of establishing bounds to the photon mass, the study 
of the Proca field equation has been a subject of constant interest in theoretical physics 
\cite{CSW_2012,DPS20015,T_2010,BCFH_2005,VIG_2002,TW97}. Proca-like 
modifications of electromagnetism 
have been considered in order to explore possible violations of Lorentz invariance at large distances
\cite{DPS20015}. The Proca equation with a negative square-mass constitutes a theoretical tool 
for the analysis of tachyon physics \cite{T_2010}. Among other interesting aspects of the 
Proca field we can mention the rich variety of phenomena
associated with the coupled Einstein-Proca field equations \cite{VIG_2002}, which constitute 
a natural extension of the celebrated Einstein-Maxwell equations. Last, but certainly not least, 
the Proca field may be related to dark matter \cite{TW97}, whose nature is one of 
the most pressing open problems in contemporary Science.

The nonlinear Schroedinger, Dirac, and Klein-Gordon equations investigated in
\cite{NRT2011,NRT2012,PT2013,CPP2013,PSNT2014,ARSML2015,RN2013} share the
physically appealing property of admitting (in the case of 
vanishing interactions) exact soliton-like localized solutions 
that behave as free particles, in the sense of complying 
with the celebrated Einstein-Planck-de Broglie relations 
connecting frequency and wave number respectively, with energy and momentum.
Given these previous developments, it is natural to ask if the
corresponding nonlinear extension can also be implemented for
massive vector bosons. This is the question we are going to explore
in the present contribution. Of especial relevance for our purposes 
is the nonlinear Klein-Gordon equation proposed in \cite{NRT2011}. It exhibits
a nonlinearity in the mass term which is proportional to a power of the
wave function $\Phi(x,t)$.  For the above mentioned exact solutions 
the wave function $\Phi(x,t)$ depends on space and time only through the combination
$x-vt$. This space-time dependence corresponds to a uniform translation at a constant
velocity $v$ without change in the shape of the wave function. These
soliton-like solutions are called  $q$-plane waves and, as already mentioned, are
compatible with the Einstein-Planck-de Broglie relations $E=\hbar w$ and $p=\hbar k$, 
satisfying the relativistic energy-momentum relation $E^2=  c^2
p^2 + m^2 c^4$. The  $q$-plane waves constitute a generalization of
the standard exponential plane waves that arise naturally within 
a theoretical framework that extends the Boltzmann-Gibbs (BG) 
entropy and statistical mechanics on the basis of a  power-law
entropic functional $S_q$. This functional is parameterized 
by a real index $q$, the usual BG formalism being recovered 
in the limit $q \rightarrow 1$. The $q$-plane waves 
are complex-valued versions of the $q$-exponential distributions
that optimize the $S_q$ entropies under appropriate constraints.
These distributions are at the core of the alluded extension
of the BG thermostatistics, which has been applied in recent years to 
a variegated set of physical scenarios. In particular, several
applications to problems in nuclear and particle physics,
as well as in quantum field theory, have been recently advanced. 
As examples we can mention applications of the $q$-nonextensive 
thermostatistics to the study of the nuclear equation of state 
\cite{SPSA2014,PSA2007,LP2012}, to neutron stars \cite{MDMC2015,LP2011}, 
to the thermodynamics of hadron systems 
\cite{P2015,D2014,MAD2013,LPQ2010},  to proton-proton and heavy ion collisions 
\cite{,AC2015,RW2014,W2014,CW2012,B2015,CHLW2009,ALQ2000}, 
to quantum cromodynamics \cite{WWCT2015,RW2009,CM2008}, to cosmic rays \cite{B2003}, 
to the Thomas-Fermi model of self-gravitating systems \cite{OT2014}, 
to fractal deformations of quantum statistics \cite{U2012},
and to the entropic-force approach to gravitation \cite{AN2014,AN2013}.
Intriguing connections between the non extensive thermostatistical
formalism and $q$-deformed dynamics have been suggested \cite{LSN2006}.
The $q$-nonextensive thermostatistics has also stimulated 
the exploration of other non-additive entropic functionals 
that have been applied, for instance, to black hole 
thermodynamics \cite{TC2013,KK2013}. 

The paper is organized as follows. In Section II we introduce a
Lagrangian for the generalized Proca field that leads to the nonlinear
Proca field equations. We study some of its main properties, with 
special emphasis on the $q$-plane wave soliton-like solutions. 
In Section III we consider the limit of massless particles, 
obtaining a new family of wave-packet solutions of Maxwell
equations. Some conclusions and final remarks are given in 
Section IV.

\section{Lagrangian Approach for Nonlinear Proca Equations}

Let us  introduce the four-dimensional space-time
operators~\cite{gregre},

\begin{equation}
\label{eq:fourdoper}
\partial^{\mu} \equiv {\partial \over \partial x_{\mu}}
\equiv \left\{
{\partial \over \partial (ct)}, -{\partial \over \partial x},
-{\partial \over \partial y}, -{\partial \over \partial z} \right\}~;
\qquad \partial_{\mu} \equiv {\partial \over \partial x^{\mu}}
\equiv \left\{
{\partial \over \partial (ct)}, {\partial \over \partial x},
{\partial \over \partial y}, {\partial \over \partial z} \right\}~,
\end{equation}

\vskip \baselineskip
\noindent
as well as the contravariant and covariant vectors

\begin{eqnarray}
\label{eq:fourdpsiphi}
&&\Psi^{\mu} \equiv (\Psi_{0}, \Psi_{x}, \Psi_{y}, \Psi_{z}); \qquad
\Psi_{\mu} \equiv (\Psi_{0}, -\Psi_{x}, -\Psi_{y}, -\Psi_{z}), \\
&&\Phi^{\mu} \equiv (\Phi_{0}, \Phi_{x}, \Phi_{y}, \Phi_{z}); \qquad
\ \Phi_{\mu} \equiv (\Phi_{0}, -\Phi_{x}, -\Phi_{y}, -\Phi_{z}).
\end{eqnarray}

\vskip \baselineskip
\noindent
As it will be shown below, these two vector fields are necessary
for a consistent field theory,
similarly to the recent nonlinear versions of the Schroedinger and
Klein-Gordon equations~\cite{NRT2012,RN2013}. For this,
we introduce the following Lagrangian density,

\begin{eqnarray}
\label{eq:genlagrannlproca}
{\cal L} &=& A \left\{
- {1 \over 2} \, F_{\mu \nu} \tilde{F}^{\mu \nu} + q \, {m^{2} c^{2} \over \hbar^{2}}
\left( \Phi_{\mu} \Psi^{\mu} \right) \left( \Psi_{\nu} \Psi^{\nu}
\right)^{q-1} \right. \nonumber \\
&-&  \left. {1 \over 2} \, F_{\mu \nu}^{*} \tilde{F}^{\mu \nu *}
+ q \, {m^{2} c^{2} \over \hbar^{2}}
\left( \Phi_{\mu}^{*} \Psi^{\mu *} \right) \left( \Psi_{\nu}^{*}
\Psi^{\nu *} \right)^{q-1}
\right\},
\end{eqnarray}

\vskip \baselineskip
\noindent
where $A$ is a multiplicative factor that may depend on the total energy
and volume (in the case the fields are confined in a finite volume $V$).
Moreover, we are adopting the
standard index-summation
convention~\cite{gregre}, and
the tensors above are given by

\begin{equation}
\label{eq:tensorfmunu}
F_{\mu \nu} = \partial_{\mu} \Phi_{\nu} - \partial_{\nu} \Phi_{\mu}~;
\qquad
\tilde{F}^{\mu \nu} = \partial^{\mu} \Psi^{\nu} - \partial^{\nu} \Psi^{\mu}~.
\end{equation}

\vskip \baselineskip
\noindent

The Euler-Lagrange equations for the vector field
$\Phi_{\mu}$~\cite{gregre},

\begin{equation}
\label{eq:eulerlagphi}
\frac{\partial {\cal L}}{\partial \Phi_{\mu}}
- \partial_{\nu} \left[ \frac{\partial {\cal L}}
{\partial (\partial_{\nu} \Phi_{\mu})} \right] = 0~,
\end{equation}

\vskip \baselineskip
\noindent
lead to

\begin{equation}
\label{eq:nlprocapsieq}
\nabla^{2} \Psi^{\mu} =
{1 \over c^{2}}
{\partial ^{2} \Psi^{\mu} \over \partial t^{2}} +
q \, {m^{2}c^{2} \over \hbar^{2}} \, \Psi^{\mu}
\left( \Psi_{\nu} \Psi^{\nu} \right)^{q-1}~,
\end{equation}

\vskip \baselineskip
\noindent
where we have used the Lorentz condition
$\partial_{\nu}\Psi^{\nu}=0$~\cite{J_1998,gregre}.
In a similar way, the Euler-Lagrange equations for the vector field
$\Psi_{\mu}$

\begin{equation}
\label{eq:eulerlagpsi}
\frac{\partial {\cal L}}{\partial \Psi_{\mu}}
- \partial_{\nu} \left[ \frac{\partial {\cal L}}
{\partial (\partial_{\nu} \Psi_{\mu})} \right] = 0~,
\end{equation}

\vskip \baselineskip
\noindent
yield

\begin{equation}
\label{eq:nlprocaphieq}
\nabla^{2} \Phi_{\mu} =
{1 \over c^{2}}
{\partial ^{2} \Phi_{\mu} \over \partial t^{2}} +
q \, {m^{2}c^{2} \over \hbar^{2}} \, \left[ \Phi_{\mu}
\left( \Psi_{\nu} \Psi^{\nu} \right)^{q-1}
+2(q-1) \Psi_{\mu} \left( \Phi_{\nu} \Psi^{\nu} \right)
\left( \Psi_{\nu} \Psi^{\nu} \right)^{q-2} \right]~.
\end{equation}

\vskip \baselineskip
\noindent
One should notice that the equations above recover the linear
Proca equations~\cite{P_1936,gregre,J_1998} in the particular
limit $q=1$, in which case $\Phi^{\mu}(\vec{x},t)=\Psi^{\mu *}(\vec{x},t)$.
Furthermore, in the one-component limit, \eq{eq:nlprocapsieq} recovers
the nonlinear Klein-Gordon equation recently proposed in
Ref.~\cite{NRT2011},

\begin{equation}
\label{eq:kgordoneq}
\nabla^{2} \Psi(\vec{x},t) =
{1 \over c^{2}}
{\partial ^{2} \Psi(\vec{x},t) \over \partial t^{2}} +
q \, {m^{2}c^{2} \over \hbar^{2}} \,
\left[ \Psi(\vec{x},t) \right]^{2q-1}~,
\end{equation}

\vskip \baselineskip
\noindent
whereas~\eq{eq:nlprocaphieq} reduces to

\begin{equation}
\label{eq:kgordoneqphi}
\nabla^{2} \Phi(\vec{x},t) =
{1 \over c^{2}}
{\partial ^{2} \Phi(\vec{x},t) \over \partial t^{2}} +
q(2q-1) \, {m^{2}c^{2} \over \hbar^{2}} \, \Phi(\vec{x},t)
\left[ \Psi(\vec{x},t) \right]^{2(q-1)}~,
\end{equation}

\vskip \baselineskip
\noindent
which corresponds to the additional nonlinear Klein-Gordon
equation, associated with the field
$\Phi(\vec{x},t)$, found by means of a Lagrangian approach in
Ref.~\cite{RN2013}.

For general $q$, the fields
$\Psi^{\mu}(\vec{x},t)$ and $\Phi^{\mu}(\vec{x},t)$ are distinct, and
the solutions of
Eqs.~(\ref{eq:nlprocapsieq}) and~(\ref{eq:nlprocaphieq}) may be written
in terms of a $q$-plane wave, similarly to the recent
nonlinear proposals of
quantum equations~\cite{NRT2011,NRT2012,RN2013}.
In fact, one has that

\begin{eqnarray}
\label{eq:nlprocapsisol}
\Psi^{\mu}(\vec{x},t) &=& a^{\mu} \exp_{q} \left[ {i \over \hbar}
(\vec{p} \cdot \vec{x} -E t) \right]~, \\
\label{eq:nlprocaphisol}
\Phi^{\mu}(\vec{x},t) &=& a^{\mu} \left\{ \exp_{q} \left[ {i \over \hbar}
(\vec{p} \cdot \vec{x} -E t) \right] \right\}^{-(2q-1)}~,
\end{eqnarray}

\vskip \baselineskip
\noindent
satisfy Eqs.~(\ref{eq:nlprocapsieq}) and~(\ref{eq:nlprocaphieq}),
provided that the coefficients are restricted
to $a_{\mu} a^{\mu}=1$.
The solutions above are expressed in terms of the
$q$-exponential function $\exp_{q}(u)$ that emerges in nonextensive
statistical mechanics~\cite{T2009}, which
generalizes the standard exponential, and for a pure imaginary $iu$, it
is defined as the principal value of

\begin{equation}
\label{eq:compqexp}
\exp_{q}(iu) = \left[ 1 + (1-q)iu \, \right]^{1 \over 1-q}; \qquad
\exp_{1}(iu) \equiv \exp(iu)~,
\end{equation}

\vskip \baselineskip
\noindent
where we used $\lim_{\epsilon \rightarrow 0} (1+\epsilon)^{1/\epsilon}=e$.
Moreover, considering these solutions,  one obtains the
energy-momentum relation

\begin{equation}
\label{eq:einsteinrel}
E^{2} = p^{2}c^{2} + m^{2}c^{4}~,
\end{equation}

\vskip \baselineskip
\noindent
from both Eqs.~(\ref{eq:nlprocapsieq}) and~(\ref{eq:nlprocaphieq}),
for all $q$. Note that, in contrast to what happens with the standard 
linear Proca equation, the four equations appearing both
in (\ref{eq:nlprocapsieq}) and in (\ref{eq:nlprocaphieq}), besides being nonlinear, 
are coupled. It is remarkable that these sets of four nonlinear coupled 
partial differential equations admit exact time dependent solutions 
of the $q$-plane wave form that are consistent with the relativistic 
energy-momentum relation (\ref{eq:einsteinrel}). 

Now, if one introduces the probability density as

\begin{equation}
\label{eq:probproca}
\rho(\vec{x},t) = {1 \over 2} \left( \Phi_{\mu} \Psi^{\mu}
+ \Phi_{\mu}^{*} \Psi^{\mu *} \right)~,
\end{equation}

\vskip \baselineskip
\noindent
the solutions of Eqs.~(\ref{eq:nlprocapsisol})
and~(\ref{eq:nlprocaphisol}) yield

\begin{equation}
\label{eq:probprocasol}
\rho(\vec{x},t) = 1 - {(1-q)^{2} \over \hbar^{2}} (\vec{p}
\cdot \vec{x} -E t)^{2}~,
\end{equation}

\vskip \baselineskip
\noindent
which require, for positiveness,

\begin{equation}
\label{eq:probprocacond}
|(1-q)(\vec{p} \cdot \vec{x} -E t)| < \hbar~.
\end{equation}

\section{Massless Particles: $q$-Plane Waves as Solutions of Maxwell Equations}

The linear Proca equations are usually considered as appropriate
for describing vectorial bosons, or massive
photons~\cite{gregre,J_1998}.
Now, if one considers the limit $m \rightarrow 0$ in the Lagrangian
of~\eq{eq:genlagrannlproca}, one eliminates its nonlinear contributions,
recovering the electromagnetic Lagrangian without sources~\cite{J_1998}.
In this case, both Eqs.~(\ref{eq:nlprocapsieq})
and~(\ref{eq:nlprocaphieq}) reduce to the standard linear wave equation,
described in terms of a single vector field (i.e.,
$\Phi^{\mu}(\vec{x},t)=\Psi^{\mu *}(\vec{x},t)$),

\begin{equation}
\label{eq:waveeq}
\nabla^{2} \Psi^{\mu} =
{1 \over c^{2}}
{\partial ^{2} \Psi^{\mu} \over \partial t^{2}}~.
\end{equation}

\vskip \baselineskip
\noindent
As usual, considering $\vec{p}=\hbar \vec{k}$ and $E=\hbar \omega$,
one verifies
easily that any twice-differentiable function of the
type $f(\vec{k} \cdot \vec{x} -\omega t)$ is a solution of the
wave equation.
In what follows we will explore the
$q$-plane wave of Eq.~(\ref{eq:nlprocapsisol}) as such a solution;
our analysis is based on
some properties of this solution, which are relevant from
the physical point of view:
(i) It presents an oscillatory behavior;
(ii) It is localized for certain values of $q$.
Indeed, for $q \neq 1$ the $q$-exponential $\exp_{q}(iu)$
is characterized by an
amplitude $r_{q} \neq 1$~\cite{B98},

\begin{equation}
\label{eq:propcompqexp1}
\exp_{q}(\pm iu) = \cos_{q}(u) \pm i \sin_{q}(u)~,
\end{equation}

\vspace{-5mm}
\begin{equation}
\label{eq:propcompqexp2}
\cos_{q}(u) = r_{q}(u)
\cos \left\{ {1 \over q-1} {\rm arctan}[(q-1)u] \right\}~,
\end{equation}

\vspace{-5mm}
\begin{equation}
\label{eq:propcompqexp3}
\sin_{q}(u) = r_{q}(u)
\sin \left\{ {1 \over q-1} {\rm arctan}[(q-1)u] \right\}~,
\end{equation}

\vspace{-5mm}
\begin{equation}
\label{eq:propcompqexp4}
r_{q}(u) = \left[1+(1-q)^{2}u^{2} \right]^{1/[2(1-q)]}~,
\end{equation}

\vskip \baselineskip
\noindent
so that  $r_{q}(u)$ decreases for
increasing arguments, if $q>1$. From
Eqs.~(\ref{eq:propcompqexp1})--(\ref{eq:propcompqexp4}) one notices
that $\cos_{q}(u)$ and $\sin_{q}(u)$
can not be zero simultaneously, yielding $\exp_{q}(\pm iu) \neq 0$;
moreover, $\exp_{q}(iu)$ presents further peculiar properties,

\begin{equation}
\label{eq:conjqexp1}
[\exp_{q}(iu)]^{*} = \exp_{q}(-iu) = \left[ 1 - (1-q)iu \,
\right]^{1 \over 1-q},
\end{equation}

\vspace{-5mm}
\begin{equation}
\label{eq:conjqexp2}
\exp_{q}(iu) [\exp_{q}(iu)]^{*} = [r_{q}(u)]^{2}
= \left[1+(1-q)^{2}u^{2} \right]^{1 \over 1-q},
\end{equation}

\vspace{-5mm}
\begin{equation}
\label{eq:sumu1u2}
\exp_{q}(iu_{1}) \exp_{q}(iu_{2})=
\exp_{q}[iu_{1}+iu_{2}-(1-q)u_{1}u_{2}]~,
\end{equation}

\vspace{-5mm}
\begin{equation}
\label{eq:conjqexp3}
\left\{ [\exp_{q}(iu)]^{\alpha}\right\}^{*} = \left\{ [\exp_{q}(iu)]^{*}
\right\}^{\alpha} = [\exp_{q}(-iu)]^{\alpha},
\end{equation}

\vskip \baselineskip
\noindent
for any $\alpha \ {\rm real}$.
By integrating~\eq{eq:conjqexp2} from $-\infty$ to $+\infty$, one
obtains~\cite{jauregui2010},

\begin{equation}
\label{eq:intrho2}
{\cal I}_{q} = \int_{-\infty}^{\infty} \ du \ [r_{q}(u)]^{2}
= \frac{\sqrt{\pi} \ \Gamma \left( \frac{3-q}{2(q-1)} \right)}
{(q-1) \ \Gamma \left({1 \over q-1}\right)}~,
\end{equation}

\vskip \baselineskip
\noindent
leading to the physically important property of square integrability
for $1<q<3$; as some simple typical examples, one
has ${\cal I}_{3/2}={\cal I}_{2}=\pi$~.
One should notice that this integral diverges in both limits
$q \rightarrow 1$ and $q \rightarrow 3$, as well as
for any $q<1$.
Hence, the $q$-plane wave of Eq.~(\ref{eq:nlprocapsisol}) presents
a modulation, typical of a localized wave, for $1<q<3$.

Then, identifying the components of the vector of~\eq{eq:fourdpsiphi}
with the scalar and vector potentials,
$\Psi^{\mu} \equiv (\phi, \vec{A})$, from which one obtains the
fields~\cite{J_1998},

\begin{equation}
\label{eq:eandbfields}
\vec{E} = -\vec{\nabla} \phi - {1 \over c} \,
{\partial \vec{A} \over \partial
t}~; \qquad \vec{B} = \vec{\nabla} \times \vec{A}~,
\end{equation}

\vskip \baselineskip
\noindent
one verifies that the wave equations for these potentials
are equivalent to the Maxwell equations in the absence of sources.

Hence, we now consider the following solutions for each Cartesian
component $j$ $(j=x,y,z)$ of the electromagnetic fields,

\begin{equation}
\label{eq:elmagfieldscomp}
E_{j}(\vec{x},t) = E_{0j} \exp_{q} \left[ i
(\vec{k} \cdot \vec{x} - \omega t) \right]~;  \quad
B_{j}(\vec{x},t) = B_{0j} \exp_{q} \left[ i
(\vec{k} \cdot \vec{x} - \omega t) \right]~,
\end{equation}

\vskip \baselineskip
\noindent
which satisfy the wave equation, for each component, provided that
$\omega = c|\vec{k}|$.
Writing the wave vector as $\vec{k}=k \vec{n}$, where $\vec{n}$ represents
a unit vector along the wave-propagation direction, the Maxwell equations
associated with the divergence of the fields $\vec{E}$ and $\vec{B}$ yield
respectively,

\begin{equation}
\label{eq:transvcond}
\vec{n} \cdot \vec{E}_{0}=0~; \quad
\vec{n} \cdot \vec{B}_{0}=0~,
\end{equation}

\vskip \baselineskip
\noindent
implying that the fields $\vec{E}$ and $\vec{B}$ are both perpendicular to
the direction of propagation.
In addition to this, considering the solutions of~\eq{eq:elmagfieldscomp}
in Faraday's Law,

\begin{equation}
\label{eq:faradaylaw}
\vec{\nabla} \times \vec{E} + {1 \over c}
\, {\partial \vec{B} \over \partial t} = 0 \quad
\Rightarrow \quad \vec{B}_{0} = \vec{n} \times \vec{E}_{0}~,
\end{equation}

\vskip \baselineskip
\noindent
whereas doing the same in Amp\`ere's Law,

\begin{equation}
\label{eq:amperelaw}
\vec{\nabla} \times \vec{B} - {1 \over c}
\, {\partial \vec{E} \over \partial t} = 0 \quad
\Rightarrow \quad \vec{E}_{0} = -\vec{n} \times \vec{B}_{0}~.
\end{equation}

\vskip \baselineskip
\noindent
The results above show that an electromagnetic wave defined
by~\eq{eq:elmagfieldscomp} corresponds to a transverse wave, similarly
to the plane-wave solution.

Now, let us consider the Poynting vector,

\begin{equation}
\label{eq:poyntingvec}
\vec{S} = {1 \over 2} {c \over 4 \pi}
\left(\vec{E} \times \vec{B}^{*} + \vec{E}^{*} \times \vec{B} \right)
= {c \over 4 \pi} |\vec{E}_{0}|^{2}
\left[1+(1-q)^{2}(\vec{k} \cdot \vec{x} - \omega t)^{2}
\right]^{1 \over 1-q} \vec{n}~,
\end{equation}

\vskip \baselineskip
\noindent
as well as the energy density,

\begin{equation}
\label{eq:energydens}
u(\vec{x},t) = {1 \over 16 \pi}
\left(\vec{E} \cdot \vec{E}^{*} + \vec{B} \cdot \vec{B}^{*} \right)
= {1 \over 8 \pi} |\vec{E}_{0}|^{2}
\left[1+(1-q)^{2}(\vec{k} \cdot \vec{x} - \omega t)^{2}
\right]^{1 \over 1-q}~.
\end{equation}

\vskip \baselineskip
\noindent
One sees from both expressions above an important difference with
respect to those associated with the standard plane-wave
solution~\cite{J_1998}, given by a factor, which is essentially
a $q$-Gaussian of the argument $|\vec{k} \cdot \vec{x} - \omega t|$.
As a consequence of this factor, one has that

\begin{equation}
\label{eq:derenergydens}
\left({\partial u \over \partial t}\right)_{\vec{x}} = {1 \over 4 \pi}|\vec{E}_{0}|^{2}
\left[1+(1-q)^{2}(\vec{k} \cdot \vec{x} - \omega t)^{2}
\right]^{q \over 1-q}(q-1) \omega
(\vec{k} \cdot \vec{x} - \omega t)~,
\end{equation}

\vskip \baselineskip
\noindent
which leads to the interesting result, $\left({\partial u / \partial t}\right)_{\vec{x}}<0$,  
for a $q$-plane wave with $q>1$, if
$\vec{k} \cdot \vec{x} < \omega t$.
This result is
directly related with the fact that the amplitude
of the wave decreases for increasing arguments, for $q>1$,
according to Eqs.~(\ref{eq:propcompqexp1})--(\ref{eq:propcompqexp4}).
Notice that
$\left({\partial u / \partial t}\right)_{\vec{x}}=0$ for $q=1$, as a consequence of 
the fact that the standard plane wave fills the whole space. 

\subsection{Physical Application: A $q$-Plane Wave in a Waveguide} 

Let us now consider the propagation of a $q$-plane wave in an infinite 
rectangular waveguide, adjusted appropriately along the wave-propagation direction, 
which will be chosen herein to be the $\vec{x}$-axis, 
i.e., $\vec{k} \cdot \vec{x}=k_{x}x$. The total energy carried by the $q$-plane
wave can be calculated from~\eq{eq:energydens}, 

\begin{equation}
\label{eq:totalenergy1}
U=\int d\vec{x} \, u(\vec{x},t) = {\sigma \over 8 \pi} |\vec{E}_{0}|^{2} \int_{-\infty}^{\infty} dx 
\left[1+(1-q)^{2}(k_{x}x - \omega t)^{2} \right]^{1 \over 1-q}~,
\end{equation}

\vskip \baselineskip
\noindent
where $\sigma$ represents the area of the transverse section of the waveguide.  
The integral above may be calculated by means of a change of variables,
$v=k_{x}x - \omega t$, in such a way to use~\eq{eq:intrho2},  

\begin{equation}
\label{eq:totalenergy2}
U = {\sigma \over 8 \sqrt{\pi}} {|\vec{E}_{0}|^{2} \over (q-1)k_{x}}  
\frac{\Gamma \left( \frac{3-q}{2(q-1)} \right)}
{\Gamma \left({1 \over q-1}\right)}~, 
\end{equation}

\vskip \baselineskip
\noindent
 leading to a finite total energy for $1<q<3$, diverging in the limit 
$q \rightarrow 1$. As a typical particular case, one has 
$U=\sigma |\vec{E}_{0}|^{2}/(8k_{x})$, for $q=2$. 
Hence, due to its localization in time, the 
total energy that a detector can absorb from the $q$-plane is finite,
in contrast to what happens with the standard plane wave.  
This enables the approach of nonlinear
excitations which do not deform in time and should be relevant, e.g.,
in nonlinear optics and plasma physics.

\section{Conclusions}

 We proposed a generalized Lagrangian that leads to a nonlinear
 extension of the Proca field equation. We discussed some of the
 main features of this nonlinear field theory, focusing on the
 existence of exact time dependent, localized solutions of the
 $q$-plane wave form. These solutions exhibit soliton-like properties, 
 in the sense of propagating with constant velocity and without changing shape. They
 have a $q$-plane wave form, which is a generalization of the standard
 complex exponential plane wave solutions of the linear Proca equation.
 The $q$-plane waves have properties suggesting that they describe free
 particles of a finite mass $m$: they are compatible with the 
 celebrated Einstein-Planck-de Broglie connection between frequency, 
 wave number, energy, and momentum, satisfying the relativistic relation
 $E = p^2 c^2 + m^2 c^4$.  
 
In the limit $q \to 1$, the present nonlinear Proca 
field theory reduces to the Maxwell linear 
electrodynamics. We see then that the $q$-deformation
associated with the nonextensive formalism turns Proca's
linear field theory into its nonlinear generalization, 
but leaves Maxwell electrodynamics invariant. As already mentioned,
for each value $q \ne 1$ the associated nonlinear Proca equation
admits $q$-plane wave solutions. In the $q \to 1$ limit, however, 
the $q$-plane wave solutions are solutions of Maxwell equations for all
$q$. 
 
The nonlinear Proca equation here introduced, together with the
nonlinear Dirac and Klein-Gordon equations previously advanced in \cite{NRT2011,RN2013}, 
provide the main ingredients of a nonlinear generalization of the 
basic relativistic field equations for particle dynamics inspired in
the $q$-thermostatistical formalism. These equations share a family 
of exact time dependent solutions: the $q$-plane waves. 
The present discussion suggests several possible directions
of future research, such as to study more complex wave-packet solutions,
and to consider interactions. Some progress in these directions has been
achieved, in a non-relativistic setting, in the case of a nonlinear
Schroedinger equation with a power-law nonlinearity in the kinetic 
energy term \cite{PT2013,CPP2013,PSNT2014}. It would be interesting 
to extend these results to relativistic scenarios. Another relevant issue 
for future exploration concerns gauge invariance. As happens with the standard
linear Proca equation, the present nonlinear Proca equation is not gauge
invariant, due to the presence of the mass term. It would be interesting to explore
whether this symmetry can be restored by adding more dynamics to the nonlinear 
Proca field theory, as is done in the liner theory via the Stueckelberg procedure \cite{GN2010}.  
Any new developments along these lines will be very welcome.

%\vskip 2\baselineskip
%\newpage

%{\large\bf Acknowledgments}

\vskip \baselineskip
%\noindent
We thank C. Tsallis for fruitful conversations.
The partial financial support from
CNPq through grant 401512/2014-2 
and from FAPERJ (Brazilian agencies) is 
acknowledged.

%\newpage

\end{document}